\documentclass[seceq]{ptptex}
\usepackage{graphicx} 
\usepackage{latexsym}
\notypesetlogo
\begin{document}
\begin{flushright} 
RUP-19-16 \\
December, 2019
\end{flushright}
\vspace{10mm}
%\begin{document}
\begin{center}
\Large{ \bf Quarks mass function at finite density \\ in real-time formalism }
\vspace{15mm}
   
\large{ Hidekazu {\sc Tanaka} and Shuji {\sc Sasagawa} \\
Department of Physics, Rikkyo University, Tokyo 171-8501, Japan\\
}
 \end{center}

\begin{center}

\vspace{25mm}

{\Large ABSTRACT}
 \end{center}
        
\vspace{10mm}
%\
%\markboth{H. {\sc Tanaka}   }{}
\def\proj{{\bf P}}  
\def\slsh#1{{#1}{\kern-6pt}/{\kern1pt}}  
%\title{ }

%\vspace{15mm}

%\author{Hidekazu {\sc Tanaka}}
%\inst{Department of Physics, Rikkyo University, \\
%           Nishi-ikebukuro, Toshima-ku Tokyo, Japan, 171 \\
%     }
%\recdate{ }
%  \begin{center}
%\maketitle
%\vspace{25mm}

%  {\Large ABSTRACT}
%  \end{center}
        
%\vspace{10mm}

Chiral symmetry restoration of quarks is investigated at finite density in quantum chromodynamics. The effective quark mass is calculated with the Schwinger-Dyson equation in the real-time formalism without the instantaneous exchange approximation. We present some properties of the quark mass functions and the quark  
propagator at zero temperature.   

% \maketitle
  \newpage

\section{Introduction}

 Evaluation of chiral phase transitions at finite density in quantum chromodynamics (QCD) is difficult task.  In order to study the chiral phase transitions, one of useful tools is the Schwinger-Dyson equation (SDE)[1,2], which can evaluate nonperturbative phenomena.

In the previous papers[3,4,5], we formulated the SDE in the real-time formalism (RTF) for QED and QCD without  the instantaneous exchange approximation (IEA)[6]. The RTF, which is formulated in Minkowski space, can evaluate non-equilibrium systems. In our method, the resonance contributions in momentum integration in Minkowski space are efficiently evaluated.

 In Ref.[5], we found that the critical temperature $T_{\rm C}$, in which the chiral symmetry is restored at zero chemical potential, is $T_{\rm C}\simeq \Lambda_{\rm QCD}/2$. Furthermore, the effective quark mass evaluated at the resonance peak of an effective quark propagator is given as $M_{\rm q} \simeq \Lambda_{\rm QCD}$.  Here $\Lambda_{\rm QCD}$ denotes the QCD scale parameter. 
  Therefore, $\Lambda_{\rm QCD}\simeq 0.32{\rm GeV}$ gives reasonable result for the effective quark mass as well as the critical temperature for the chiral phase transition.

 In this paper, we study properties of the quark mass function with the SDE in the RTF beyond the IEA for finite density at zero temperature, which corresponds to a high density matter at low temperature, such as internal structure of neutron stars. 

In section 2, we present formula for the SDE in the RTF without the IEA. In section 3, some numerical results for the effective quark mass are calculated. In order to investigate instability of the massive quark state, we evaluate time dependences of the effective quark propagator.
 Section 4 is devoted to the summary and some comments. 

\section{SDE for quark mass function}

In the RTF, we implement two types of fields specified by 1 and 2 in the theory. The type-1 field is the usual field and the type-2 field corresponds to a ghost filed in the heat bath.
 
In one-loop order, we calculate the 1-1 component of a self-energy of quark $\Sigma^{11}(P)$ in QCD , which is given by
\begin{eqnarray}
 -i\Sigma^{11}(P)=(ig_{\rm s})^2C_F\int{d^4q \over (2\pi)^4}\gamma^{\mu}iS^{11}(Q)\Gamma^{\nu}iD^{11}_{\mu\nu}(K),
\end{eqnarray}
where $S^{11}(Q)$ and $D_{\mu\nu}^{11}(K)$ are the 1-1 components of thermal propagators for a quark with momentum $Q=(q_0,{\bf q})$ and a gluon with momentum $K=P-Q=(k_0,{\bf k})$, respectively. An external momentum of the quark is denoted by $P=(p_0,{\bf p})$. In our formulation, the time evolution of the system is generated with  an operator ${\hat H}'={\hat H}-\mu{\hat N}$.  Here ${\hat H}$ and ${\hat N}$ denote a hamiltonian and a number operator of the quark, respectively. The energy eigenvalues for ${\hat H}'$ are denoted by $p_0$ and $q_0$.
In our calculation, the quark-gluon vertex is defined by $\Gamma^{\nu}=\gamma^{\nu}$ with the gamma matrices $\gamma^{\nu}$. 
  The strong coupling constant and the color factor are denoted by $g_{\rm s}$ and $C_F=4/3$, respectively.

The 1-1 component of the quark propagator in the RTF is given as
\begin{eqnarray}
iS^{11}(Q)  = i\left[(S_{\rm F}(Q))_{\rm R}+i(S_{\rm F}(Q))_{\rm I}N_{\rm F}(\mu,q_0) \right], 
\end{eqnarray}
where $N_{\rm F}(\mu,q_0)=\epsilon(q_0+\mu)\epsilon(q_0)$ with a chemical potential $\mu$ at zero temperature, in which we define $\epsilon(z)=\theta(z)-\theta(-z)$ with the step function $\theta(z)$.
Here, $(S_{\rm F}(Q))_{\rm R}$ and $(S_{\rm F}(Q))_{\rm I}$ are the real and imaginary parts of the quark propagator $S_{\rm F}(Q)$, respectively. The quark propagator $S_{\rm F}(Q)$ is defined as
\begin{eqnarray}
iS_{\rm F}(Q)  \equiv i(\slsh{Q}+\gamma^0\mu+M(Q)) I_{\rm F}(Q).
\end{eqnarray}
with
\begin{eqnarray}
I_{\rm F}(Q) = {1 \over (q_0+\mu)^2-{\bf q}^2-M^2(Q)+i\varepsilon}.
\end{eqnarray}

The 1-1 component of the gluon propagator is given as 
 \begin{eqnarray}
iD_{\mu\nu}^{11}(K)= iD_{{\rm F}\mu\nu}(K)=i\left[(D_{{\rm F}\mu\nu}(K))_{\rm R}+i(D_{{\rm F}\mu\nu}(K))_{\rm I}\right], 
\end{eqnarray}
where
 \begin{eqnarray}
iD_{{\rm F}\mu\nu}(K)\equiv P^{\rm L}_{\mu\nu}iD_{\rm L}(K)+P^{\rm T}_{\mu\nu}iD_{\rm T}(K)
\end{eqnarray}
with
 \begin{eqnarray}
P^{\rm L}_{\mu\nu}=-g_{\mu\nu}+{K_{\mu}K_{\nu} \over K^2}-P^{\rm T}_{\mu\nu}
\end{eqnarray}
and
 \begin{eqnarray}
P^{\rm T}_{\mu\nu}=\left(-g_{\mu\nu}+{K_{\mu}K_{\nu}\over {\bf k}^2}\right)(1-\delta_{0\mu})(1-\delta_{0\nu}),
\end{eqnarray}
where the longitudinal and transverse components of the gluon propagator are given as  
\begin{eqnarray}
iD_{\rm L}(K)={i \over K^2- m_{\rm L}^2+i\varepsilon} 
\end{eqnarray}
and 
 \begin{eqnarray}
iD_{\rm T}(K)={i \over K^2- m_{\rm T}^2+i\varepsilon}, 
\end{eqnarray}
respectively. Here, $m_{\rm L}$ and $m_{\rm T}$ denote the longitudinal and transverse gluon masses, respectively. 

Integrating over the azimuthal angle of the quark momentum ${\bf q}$,  the trace of the self-energy $\Sigma^{11}$ is given by
\begin{eqnarray}
M^{11}(P) \equiv{1 \over 4}Tr[\Sigma^{11}(P)] = -{iC_F \over 2\pi^2}\int^{\Lambda_0}_{-\Lambda_0}dq_0 \int^{\Lambda}_{\delta} dq {q \over p} \alpha_{\rm s} [MI_{11}J_{11}](P,Q) 
\end{eqnarray}
with $p=|{\bf p}|$,$q=|{\bf q}|$, and $\alpha_{\rm s}=g_{\rm s}^2/(4\pi)$, \footnote{ In numerical calculations, the strong coupling constant $\alpha_{\rm s}$ is replaced by the running coupling constant $\alpha_{\rm s}(t) = g^2_{\rm s}(t)/( 4\pi) $[7] with $t=\log[({\bar P}^2+{\bar Q}^2+\mu^2)/\Lambda^2_{\rm QCD}]$,where ${\bar P}^2=p_0^2+p^2$ and ${\bar Q}^2=q_0^2+q^2$.
}
 where,
\begin{eqnarray}
MI_{11}= (MI_{\rm F})_R+i(MI_{\rm F})_{\rm I}N_{\rm F}(\mu,q_0)
\end{eqnarray}
and
\begin{eqnarray}
J_{11}=(J_{\rm F})_{\rm R}+i(J_{\rm F})_{\rm I}. 
\end{eqnarray}
with
\begin{eqnarray}
J_{\rm F}=\int^{\eta_+}_{\eta_-}dkk\left[D_{\rm L}(K)+2D_{\rm T}(K)\right]
\end{eqnarray}
with $\eta_{\pm}=|p\pm q|$ and $k=|{\bf k}|$, respectively.

The real part $M_{\rm R} $ and  the imaginary part $M_{\rm I} $ of the mass $M$ in Eq.(2$\cdot$3) are given by $M_{\rm R}=(M^{11})_{\rm R}$ and $M_{\rm I}=(M^{11})_{\rm I}/N_{\rm F}(\mu,q_0)$,respectively.  On the other hand, the real part $(M^2)_{\rm R} $ and  the imaginary part $(M^2)_{\rm I} $ of the mass $M^2$ in Eq.(2$\cdot$4) are given by $(M^2)_{\rm R}=(M_{\rm R})^2-(M_{\rm I})^2$ and $(M^2)_{\rm I}=2M_{\rm R}M_{\rm I}$,respectively.  

In Minkowski space, if the imaginary part of the mass function $(M^2)_{\rm I}$ is small, the quark propagator $I_{\rm F}$ in Eq.(2$\cdot$4)  varies rapidly near $(q_0+\mu)^2-q^2 \simeq (M^2)_{\rm R}$. As implemented in the previous works [4,5,6], we divide the $q_0$ integration into small ranges and integrate the quark propagator over $q^{(l)}_0\leq q_0 \leq q^{(l+1)}_0~(q_0^{(1)}=-\Lambda_0,q_0^{(N)}=\Lambda_0)$, in which   remaining contributions of the integrand are averaged over the range $q^{(l)}_0\leq q_0 \leq q^{(l+1)}_0$.

In order to investigate instability of the massive quark state, we evaluate time dependences of the quark propagator
\begin{eqnarray}
i{\tilde S}_{\rm F}(x_0, {\bf p})= \int d^3 x  iS_{\rm F}(x_0,{\bf x})e^{-i{\bf x}\cdot{\bf p}}
\end{eqnarray}
 with
\begin{eqnarray}
iS_{\rm F}(x_0,{\bf x})= \int {d^4 Q \over (2\pi)^4}iS_{\rm F}(Q)e^{-i(x_0q_0-{\bf x}\cdot{\bf q})},
\end{eqnarray}
where $x_0$ and ${\bf x}$ denote a time and a space coordinates, respectively. 

Integrating over ${\bf q}$, the quark propagator $ i{\tilde S}_{\rm F}(x_0, {\bf p})$ is given as 
\begin{eqnarray}
i{\tilde S}_{\rm F}(x_0, {\bf p})= \int {dq_0 \over 2\pi}iS_{\rm F}({\tilde Q})e^{-ix_0q_0}
\end{eqnarray}
with ${\tilde Q}=(q_0,{\bf p})$.
We separate $ i{\tilde S}_{\rm F}(x_0, {\bf p})$ as
\begin{eqnarray}
i{\tilde S}_{\rm F}(x_0, {\bf p})= i{\tilde S}_{\rm F}^{(+)}(x_0, {\bf p})+ i{\tilde S}_{\rm F}^{(-)}(x_0, {\bf p})
\end{eqnarray}
with
\begin{eqnarray}
i{\tilde S}_{\rm F}^{(\pm)}(x_0, {\bf p})=\pm\int {dq_0  \over 2\pi} {i(\slsh{\tilde{Q}}+\gamma^0\mu+M(\tilde{Q})) \over (q_0+\mu)\mp E(\tilde{Q}) }{1 \over 2E(\tilde{Q})}e^{-ix_0q_0},
\end{eqnarray}
 where $E$ is the quark energy. The real and imaginary parts of the quark energy $E_{\rm R}$ and $E_{\rm I} $ are defined by $ E_{\rm R}=|E|\cos(\Phi/2)$ and $ E_{\rm I}=|E|\sin(\Phi/2)$ with $\Phi=\arctan((E^2)_{\rm I}/(E^2)_{\rm R})$ and $|E|=\sqrt{|E^2|}=((E^2)_{\rm R}^2+(E^2)_{\rm I}^2)^{1/4}$. Here the real and imaginary parts of the squared quark energy $E^2$ are defined as $(E^2)_{\rm R}={\bf p}^2+(M^2)_{\rm R}$ and $(E^2)_{\rm I}=(M^2)_{\rm I}-\varepsilon $, respectively.

Here, $ i{\tilde S}_{\rm F}^{(\pm)}$ are further written by
\begin{eqnarray}
i{\tilde S}_{\rm F}^{(\pm)}(x_0, {\bf p})=\pm i\gamma_0\int {dq_0  \over 2\pi} {1 \over 2E(\tilde{Q})}e^{-ix_0q_0}+i{\bar S}_{\rm F}^{(\pm)}(x_0, {\bf p})
\end{eqnarray}
with 
\begin{eqnarray}
i{\bar S}_{\rm F}^{(\pm)}(x_0, {\bf p})=\pm \int {dq_0  \over 2\pi} {i(\slsh{\bar{Q}}_{\pm} + M(\tilde{Q})) \over (q_0+\mu)\mp E(\tilde{Q}) }{1 \over 2E(\tilde{Q})}e^{-ix_0q_0},
\end{eqnarray}
where $\bar{Q}_{\pm}=(\pm E,{\bf p})$. Here the first terms in Eq.(2$\cdot$20) are canceled in $ i{\tilde S}_{\rm F}(x_0, {\bf p})$.

Using 
\begin{eqnarray}
{1 \over (q_0+\mu)- E(\tilde{Q}) }e^{-ix_0q_0}=-i\int^{x_0}_{-\infty}dy_0e^{-iq_0y_0-i(E(\tilde{Q})-\mu)(x_0-y_0)}
\end{eqnarray}
for $E_{\rm I}<0$ and  
\begin{eqnarray}
{1 \over (q_0+\mu)- E(\tilde{Q}) }e^{-ix_0q_0}=i\int_{x_0}^{\infty}dy_0e^{-iq_0y_0-i(E(\tilde{Q})-\mu)(x_0-y_0)}
\end{eqnarray}
for $E_{\rm I}>0$, the quark propagator $ i{\bar S}_{\rm F}^{(+)}(x_0, {\bf p})$ is given as
\begin{eqnarray}
i{\bar S}_{\rm F}^{(+)}(x_0, {\bf p})=\int {dq_0 \over 2\pi} \int^{x_0}_{-\infty}dy_0{(\slsh{\bar Q}_{+}+M({\tilde Q})) \over 2E({\tilde Q}) }e^{-iy_0q_0-i(E({\tilde Q})-\mu)(x_0-y_0)}
\end{eqnarray}
for $E_{\rm I}<0$ and
\begin{eqnarray}
i{\bar S}_{\rm F}^{(+)}(x_0, {\bf p})=-\int {dq_0 \over 2\pi} \int_{x_0}^{\infty}dy_0{(\slsh{\bar Q}_{+}+M({\tilde Q})) \over 2E({\tilde Q}) }e^{-iy_0q_0-i(E({\tilde Q})-\mu)(x_0-y_0)}
\end{eqnarray}
for $E_{\rm I}>0$, respectively.

Similarly using 
\begin{eqnarray}
{1 \over (q_0+\mu)+ E(\tilde{Q}) }e^{-ix_0q_0}=i\int_{x_0}^{\infty}dy_0e^{-iq_0y_0+i(E(\tilde{Q})+\mu)(x_0-y_0)}
\end{eqnarray}
for $E_{\rm I}<0$ and  
\begin{eqnarray}
{1 \over (q_0+\mu)+ E(\tilde{Q}) }e^{-ix_0q_0}=-i\int^{x_0}_{-\infty}dy_0e^{-iq_0y_0+i(E(\tilde{Q})+\mu)(x_0-y_0)}
\end{eqnarray}
for $E_{\rm I}>0$, the quark propagator $ i{\bar S}_{\rm F}^{(-)}(x_0, {\bf p})$ is given as
\begin{eqnarray}
i{\bar S}_{\rm F}^{(-)}(x_0, {\bf p})=\int {dq_0 \over 2\pi} \int_{x_0}^{\infty}dy_0{(\slsh{\bar{Q}}_{-}+M(\tilde{Q})) \over 2E({\tilde Q}) }e^{-iy_0q_0+i(E(\tilde{Q})+\mu)(x_0-y_0)}
\end{eqnarray}
for $E_{\rm I}<0$ and
\begin{eqnarray}
i{\bar S}_{\rm F}^{(-)}(x_0, {\bf p})=-\int {dq_0 \over 2\pi} \int^{x_0}_{-\infty}dy_0{(\slsh{\bar{Q}}_{-}+M(\tilde{Q})) \over 2E(\tilde{Q}) }e^{-iy_0q_0+i(E(\tilde{Q})+\mu)(x_0-y_0)}
\end{eqnarray}
for $E_{\rm I}>0$, respectively.

\section{Numerical results}

In this section, some numerical results are presented. We solve the SDE presented in Eq. (2$\cdot$11) by a recursion method starting from a constant mass at $\mu=0$. \footnote{The initial input parameters are $M_{\rm R}=\Lambda_{\rm QCD}$ and $M_{\rm I}=0$ at $\mu=0$ with $\Lambda_0=\Lambda=10\Lambda_{\rm QCD}$ and $\delta=0.1\Lambda_{\rm QCD}$ with $\varepsilon=10^{-6}$. 
In evaluation of the quark mass function at $\mu+\Delta \mu$, we implement the solution of $M$ obtained at $\mu $  as the initial input. 
% Here, we define $M= M_{\rm R}+i M_{\rm I}$.
 }

 For each iteration, the mass function is normalized by a current quark mass at large $\zeta^2=p_0^2- p^2$, in which perturbative calculations are reliable. 

In the iteration, the mass function $M(p_0,p)$ in integrand of  the SDE is replaced by the renormalized one obtained by the previous iteration. 
\footnote{We take the renormarized mass $m(\zeta^2)=0$ at $\zeta=10\Lambda_{\rm QCD}$.  }

In this paper, we evaluate integrated mass functions $\langle|M|\rangle,\langle M_{\rm R}\rangle$ and $\langle M_{\rm I}\rangle$ as order parameters, in which $|M(p_0,p)|,M_{\rm R}(p_0,p)$ and $M_{\rm I}(p_0,p)$ are integrated over $p_0$ and $p$, respectively.[5]
\begin{figure}
\centerline{\includegraphics[width=10cm]{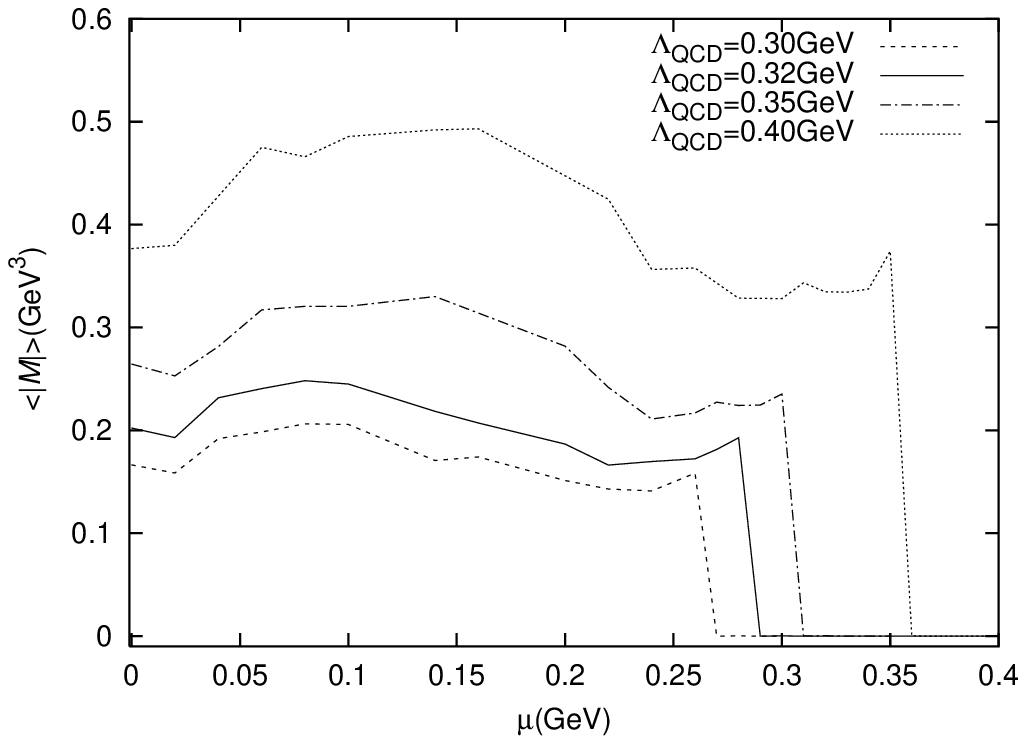}}
\caption{The $\mu$ dependences of the integrated quark mass functions $\langle|M|\rangle$ with $\Lambda_{\rm QCD}=0.30~{\rm GeV}$,$0.32~{\rm GeV}$,$0.35~{\rm GeV}$ and $0.40~{\rm GeV}$, respectively, with the massless gluon. }
%\label{Fig.1}
\end{figure}

In Fig.1, the $\mu$ dependences of $\langle|M|\rangle $ for $\Lambda_{\rm QCD}=0.30~{\rm GeV},0.32~{\rm GeV},0.35~{\rm GeV}$ and $0.40~{\rm GeV}$ with the massless gluon are presented at $T=0$.
The  transition of the chiral symmetry restoration seems to be the first order. The critical chemical potential of the phase transition  $\mu_{\rm C}$ depends on  the  QCD parameter $\Lambda_{\rm QCD}$. 
In our calculation, $0.30  ~{\rm GeV} \leq \Lambda_{\rm QCD} \leq 0.40  ~{\rm GeV}$ gives $0.27  ~{\rm GeV} \leq \mu_{C} \leq 0.36  ~{\rm GeV}$, roughly $\mu_{\rm C} \sim 0.9\Lambda_{\rm QCD}$.

 In order to choose the QCD parameter $\Lambda_{\rm QCD}$, we need another condition.
Our model roughly gives the real part of the squared quark mass function  $ (M^2)_{\rm R} \simeq \Lambda_{\rm QCD}^2$ at $\mu=T=0$.
   Here, $ (M^2)_{\rm R}$ is determined by the resonance peak of the quark propagator. 
 In our calcularion, $\Lambda_{\rm QCD}=0.32  ~{\rm GeV}$ gives $\sqrt{(M^2)_R}\simeq 0.32  ~{\rm GeV}$ at $T=\mu=0$ and the critical temperature of the chiral symmetry restoration $T_{\rm C}\simeq 0.175  ~{\rm GeV}$ with $\mu=0$. [5]

\begin{figure}
\centerline{\includegraphics[width=10cm]{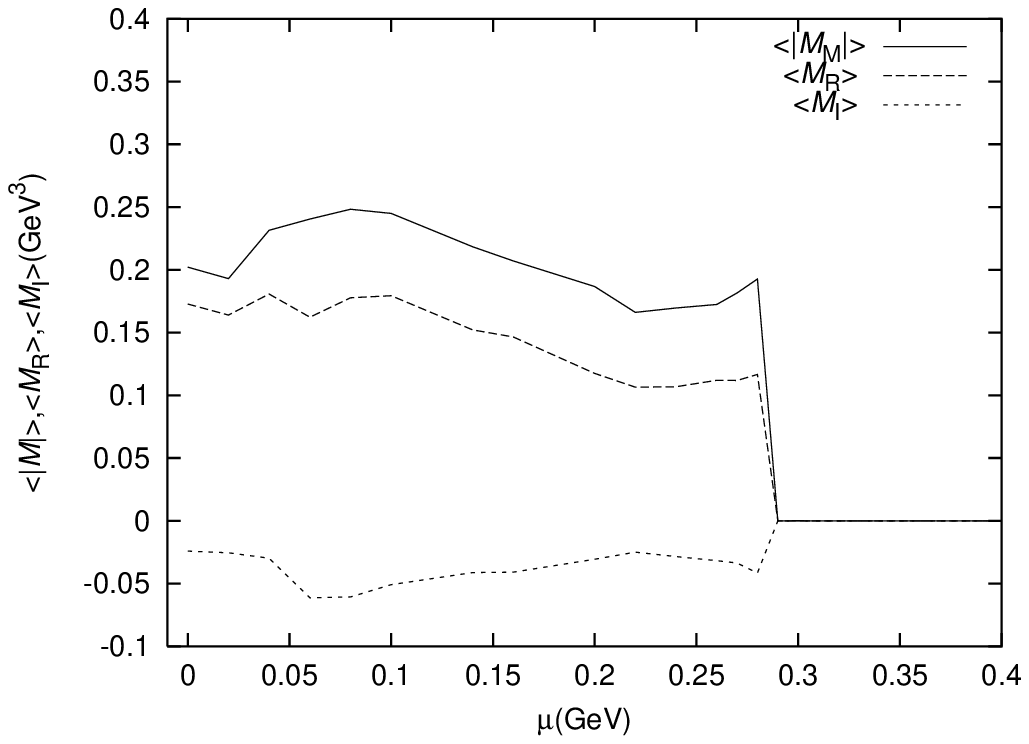}}
\caption{The $\mu$ dependences of the integrated quark mass functions $\langle|M|\rangle$,$\langle M_{\rm R}\rangle$ and $\langle M_{\rm I}\rangle$ at $\Lambda_{\rm QCD}=0.32  ~{\rm GeV}$ with the massless gluon. }
%\label{Fig.1}
\end{figure}

In Fig.2, the $\mu$ dependences of the integrated quark mass functions $\langle|M|\rangle,\langle M_R\rangle$ and $\langle M_I\rangle$ with the massless gluon are presented at $\Lambda_{\rm QCD}=0.32   ~{\rm GeV}$, which gives $\mu_{\rm C}\simeq 0.29  ~{\rm GeV}$.

As shown in Fig.2, the imaginary part of the mass function $\langle M_{\rm I}\rangle $ is non-zero value for broken chiral symmetric phase below  the critical chemical potential  $\mu_{\rm C}$, which means the massive quark state may be unstable if energy scale  rapidly changes. 
 Furthermore, the real and imaginary parts vanish at the same critical chemical potential. 
   
%The imaginary part of the mass function gives imaginary part of the quark energy, which may contribute the time evolution of the quark propagator. 
 In order to investigate instability of the massive quark state, we calculate a time dependence of the quark propagator $i{\bar S}_{\rm F}^{(+)}(x_0, {\bf p})$ in Eq.($2\cdot 21$),which is separated as
\begin{eqnarray}
i{\bar S}_{\rm F}^{(+)}(x_0, {\bf p})= \gamma_{\mu}iS_{+}^{\mu}(x_0, {\bf p})+ iS_+^{\rm M}(x_0, {\bf p})
\end{eqnarray}
with
\begin{eqnarray}
iS_{+}^{\mu}(x_0, {\bf p})= i\int {dq_0  \over 2\pi} {{\bar Q}_+^{\mu} \over (q_0+\mu)- E(q_0,p) } {1 \over 2E(q_0,p) } e^{-ix_0q_0}
\end{eqnarray}
and
\begin{eqnarray}
iS_{+}^{\rm M}(x_0, {\bf p})= i\int {dq_0  \over 2\pi} {M(q_0,p) \over (q_0+\mu)- E(q_0,p) } {1 \over 2E(q_0,p)} e^{-ix_0q_0}.
\end{eqnarray}
From Eqs.(2$\cdot$24) and (2$\cdot$25), the real and imaginary parts of $ iS_{+}^{\mu} (x_0, {\bf p})$ and $ iS_{+}^{\rm M}(x_0, {\bf p})$ are given as
\begin{eqnarray*}
{\rm Re}\left[iS_{+}^{\mu} (x_0, {\bf p})\right]=\int^{x_0}_{-x_{\rm M}}dy_0\int_{-\Lambda_0}^{\Lambda_0}{dq_0 \over 2\pi}{|{\bar Q}_+^{\mu}|\over 2|E(q_0,p)|}\cos(\Psi^{\mu}) e^{E_{\rm I}(q_0,p)(x_0-y_0)} \theta(-E_{\rm I}(q_0,p))
\end{eqnarray*}
\begin{eqnarray}
-\int_{x_0}^{ x_{\rm M}}dy_0\int_{-\Lambda_0}^{\Lambda_0}{dq_0 \over 2\pi}{|{\bar Q}_+^{\mu}|\over 2|E(q_0,p)|}\cos(\Psi^{\mu}) e^{E_{\rm I}(q_0,p)(x_0-y_0)} \theta(E_{\rm I}(q_0,p))
\end{eqnarray}
and
\begin{eqnarray*}
{\rm Im}\left[iS_{+}^{\mu} (x_0, {\bf p})\right]=-\int^{x_0}_{- x_{\rm M}}dy_0\int_{-\Lambda_0}^{\Lambda_0}{dq_0 \over 2\pi}{|{\bar Q}_+^{\mu}|\over 2|E(q_0,p)|}\sin(\Psi^{\mu}) e^{E_{\rm I}(q_0,p) (x_0-y_0)} \theta(-E_{\rm I}(q_0,p))
\end{eqnarray*}
\begin{eqnarray}
+\int_{x_0}^{ x_{\rm M}}dy_0\int_{-\Lambda_0}^{\Lambda_0}{dq_0 \over 2\pi}{|{\bar Q}_+^{\mu}|\over 2|E(q_0,p)|}\sin(\Psi^{\mu}) e^{E_{\rm I}(q_0,p) (x_0-y_0)} \theta(E_{\rm I}(q_0,p)),
\end{eqnarray}
with
\begin{eqnarray}
\Psi^{\mu}=q_0y_0+(E_{\rm R}(q_0,p)-\mu)(x_0-y_0)+\Phi_Q^{\mu}-\Phi/2
\end{eqnarray}
for $iS_{+}^{\mu} (x_0, {\bf p})$, and
\begin{eqnarray*}
{\rm Re}\left[iS_{+}^{\rm M} (x_0, {\bf p})\right]=\int^{x_0}_{-x_{\rm M}}dy_0\int_{-\Lambda_0}^{\Lambda_0}{dq_0 \over 2\pi}{|M(q_0,p)|\over 2|E(q_0,p)|}\cos(\Psi^{\rm M}) e^{E_{\rm I}(q_0,p)(x_0-y_0)} \theta(-E_{\rm I}(q_0,p))
\end{eqnarray*}
\begin{eqnarray}
-\int_{x_0}^{ x_{\rm M}}dy_0\int_{-\Lambda_0}^{\Lambda_0}{dq_0 \over 2\pi}{|M(q_0,p)|\over 2|E(q_0,p)|}\cos(\Psi^{\rm M }) e^{E_{\rm I}(q_0,p)(x_0-y_0)} \theta(E_{\rm I}(q_0,p))
\end{eqnarray}
and
\begin{eqnarray*}
{\rm Im}\left[iS_{+}^{\rm M } (x_0, {\bf p})\right]=-\int^{x_0}_{- x_{\rm M}}dy_0\int_{-\Lambda_0}^{\Lambda_0}{dq_0 \over 2\pi}{|M(q_0,p)|\over 2|E(q_0,p)|}\sin(\Psi^{\rm M }) e^{E_{\rm I}(q_0,p) (x_0-y_0)} \theta(-E_{\rm I}(q_0,p))
\end{eqnarray*}
\begin{eqnarray}
+\int_{x_0}^{ x_{\rm M}}dy_0\int_{-\Lambda_0}^{\Lambda_0}{dq_0 \over 2\pi}{|M(q_0,p)|\over 2|E(q_0,p)|}\sin(\Psi^{\rm M }) e^{E_{\rm I}(q_0,p) (x_0-y_0)} \theta(E_{\rm I}(q_0,p)),
\end{eqnarray}
with
\begin{eqnarray}
\Psi^{\rm M }=q_0y_0+(E_{\rm R}(q_0,p)-\mu)(x_0-y_0)+\Phi_{\rm M} -\Phi/2
\end{eqnarray}
for $iS_{+}^{\rm M} (x_0, {\bf p})$,respectively.\footnote{$x_{\rm M}$ is taken as $ x_{\rm M}=10x_{\rm max}$. Here, $x_{\rm max}$ is an maximum value of the plot for $x_0$.} 
Here, $\Phi_{\rm Q}^{\mu}$ and $\Phi_{\rm M}$ are defined as
\begin{eqnarray}
\Phi_{\rm Q}^0=\Phi/2
 \end{eqnarray}
and
\begin{eqnarray}
\Phi_{\rm Q}^i=\pi\theta(-p^i)
 \end{eqnarray}
for $i=1,2,3$,
and 
\begin{eqnarray}
\Phi_M=\arctan{M_{\rm I} \over M_{\rm R}},
 \end{eqnarray}
respectively.  In our calculations, we set the momentum ${\bf p}$ as ${\bf p}=(p^1,p^2,p^3)=(0,0,p)$
\begin{figure}
\centerline{\includegraphics[width=10cm]{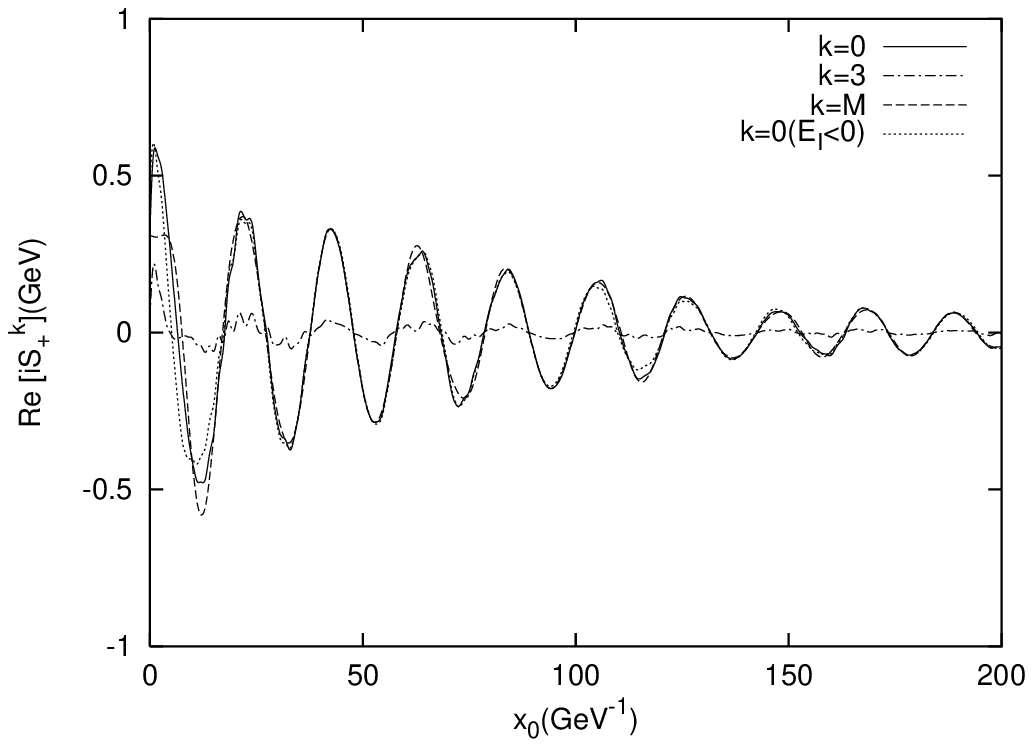}}
\caption{The $x_0$ dependences of the quark propagator ${\rm Re}\left[i S_{+}^{\rm k}(x_0,{\bf \delta})\right]$ with $\mu=0.0 ~{\rm GeV}$ and $p=\delta=0.1\Lambda_{\rm QCD}$for $k=0,~ 3,~{\rm M}$,respectively, at $\Lambda_{\rm QCD}=0.32  ~{\rm GeV}$ with the massless gluon. The dotted  curve denotes the quark propagator ${\rm Re}\left[iS_0 (x_0,{\bf \delta})\right]$ for $E_{\rm I}<0 $}
%\label{Fig.3}
\end{figure}

In Fig.3, the real part of $ iS_{+}^{\rm k} (x_0, {\bf p})$ with $p=\delta=0.1\Lambda_{\rm QCD}$ for $k=0,~ 3,~{\rm M }$,respectively, are presented at $\Lambda_{\rm QCD}=0.32  ~{\rm GeV}$ and $\mu=0.0~{\rm GeV}$ with the massless gluon. We can see that $ iS_{+}^{0} (x_0, {\bf \delta})\simeq  iS_{+}^{\rm M} (x_0, {\bf \delta}) \gg iS_{+}^{3} (x_0, {\bf \delta})$, since $|E|\simeq |M|\gg \delta$.
As shown in Fig.3, the amplitude of the quark propagator decreases as increasing $x_0$, which means the imaginary part of the quark energy $E_{\rm I}$ plays a role of decay constant. The dotted  curve denotes the quark propagator ${\rm Re}\left[iS_{+}^{\rm M} (x_0, {\bf \delta})\right]$  for $E_{\rm I}<0 $. Therefore, the contribution from $E_{\rm I}>0$ is not significant for the time evolution of the quark propagator.

\begin{figure}
\centerline{\includegraphics[width=10cm]{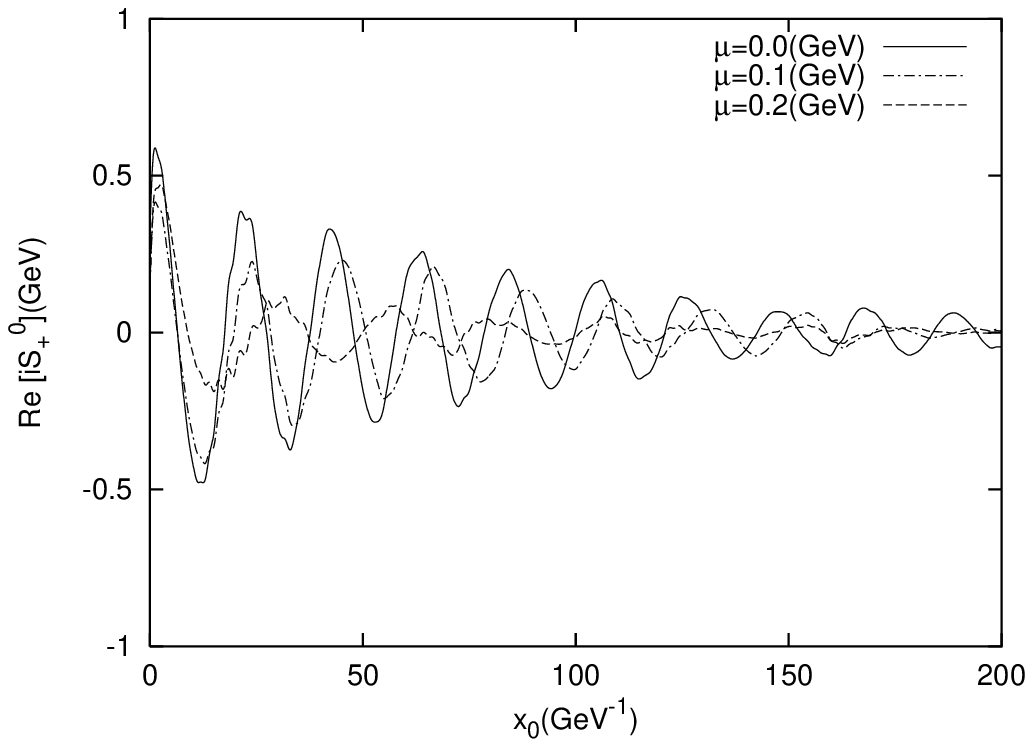}}
\caption{The $x_0$ dependences of the quark propagator ${\rm Re}\left[i S_{+}^{0}(x_0,{\bf \delta})\right]$ for $\mu=0.0  ~{\rm GeV},0.1  ~{\rm GeV}$ and $0.2  ~{\rm GeV}$,respectively, at $\Lambda_{\rm QCD}=0.32  ~{\rm GeV}$ with the massless gluon. }
%\label{Fig.4}
\end{figure}
In Fig.4, the $x_0$ dependences of the quark propagator ${\rm Re}\left[i S_{+}^{0}(x_0,{\bf \delta})\right]$ for $\mu=0.0  ~{\rm GeV},0.1  ~{\rm GeV}$ and $0.2  ~{\rm GeV}$,respectively, are presented  at $\Lambda_{\rm QCD}=0.32  ~{\rm GeV}$ with the massless gluon.
As shown in Fig.4, the wavelength is longer as $\mu$ increases due to the term $(E_{\rm R}-\mu)x_0 $ of the phase $\Psi^{0}$ in Eq.(3$\cdot$6).

\begin{figure}
\centerline{\includegraphics[width=10cm]{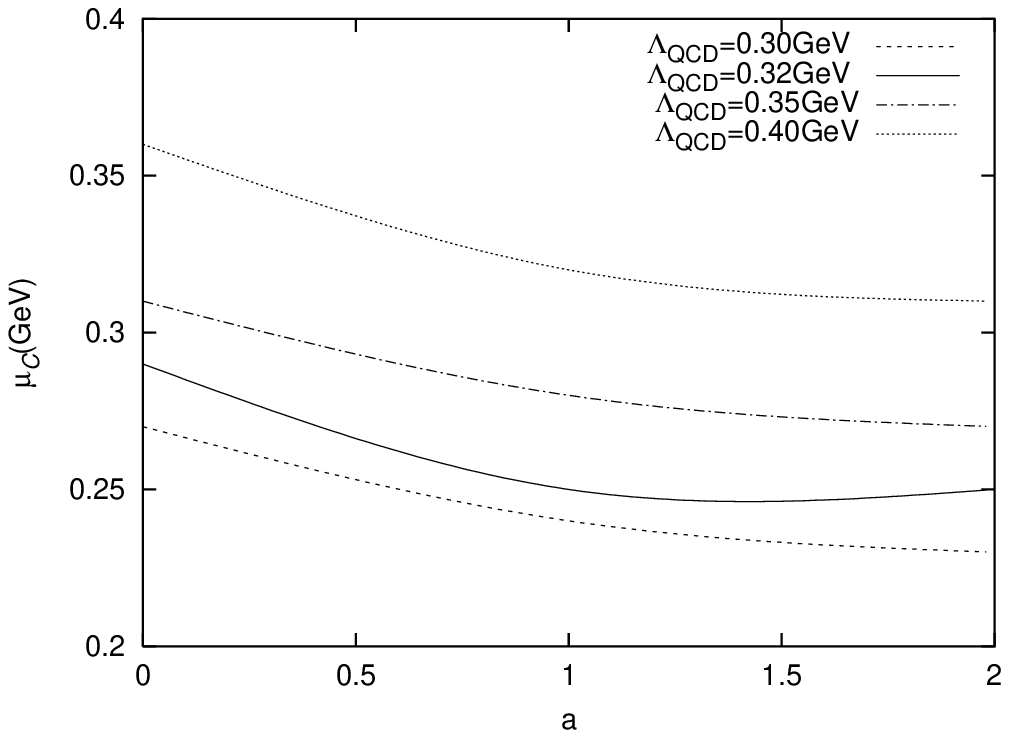}}
\caption{ The $a$ dependences on $\mu_C$ with the gluon masses defined by $m_T=0$ and $m_L=a\Lambda_{QCD}$ with $\Lambda_{\rm QCD}=0.30  ~{\rm GeV}$,$0.32  ~{\rm GeV}$,$0.35  ~{\rm GeV}$ and $0.40  ~{\rm GeV}$, respectively.
}
\end{figure}

In Fig.5, the gluon mass dependences are presented for different values of $\Lambda_{QCD}$ with  $\Lambda_{\rm QCD}=0.30  ~{\rm GeV}$,$0.32  ~{\rm GeV}$,$0.35  ~{\rm GeV}$ and $0.40  ~{\rm GeV}$, respectively.
The gluon masses are defined as $m_T=0$ and $m_L=a\Lambda_{QCD}$ with a parameter $a$.
Though the critical chemical potential $\mu_C$ depends on the gluon mass, the gluon mass dependence on  the effective quark mass  seems to be weaker for larger gluon mass.

\section{Summary and Comments}

 In this paper, we have investigated quark mass functions solved by the Schwinger-Dyson equation (SDE) at finite density with zero temperature in the real-time formalism (RTF).

 In our model, the critical chemical potential $\mu_{\rm C}$, in which the chiral symmetry is restored, depends on the QCD scale parameter $\Lambda_{\rm QCD}$ and the gluon masses ($m_{\rm L}$ and $m_{\rm T}$).   Here,$m_{\rm L}$ and $m_{\rm T}$ denote the masses for a longitudinal component and a transverse component of the gluon propagator, respectively. Our model roughly gives the critical chemical potential for the chiral symmetry restoration $\mu_{\rm C}\sim 0.9\Lambda_{\rm QCD}$ at $T=0$ with a massless gluon  ($m_{\rm L}=m_{\rm T}=0$).
The transition of the chiral symmetry restoration seems to be the first order at $T=0$.
 
We found that the imaginary part of the integrated mass function $\langle M_{\rm I}\rangle $ is non-zero value for broken chiral symmetric phase, which means  that  the massive quark state may be unstable for $\mu < \mu_{\rm C}$.
@Furthermore, the real and imaginary parts of the integrated mass functions vanish at the same critical point. 

In order to examine the effect of the imaginary part of the quark energy $E_{\rm I}$, we calculated the time evolution of the quark propagator.
 The quark propagator decreases as increasing the time, which suggests that main contribution of   the imaginary part of the energy comes from $ E_{\rm I}<0$. The contribution from $ E_{\rm I}>0$ does not give significant contribution for the time evolution of the quark propagator.  

We also calculated the gluon mass dependence of the quark mass function with a simple ansatz. We presented the critical chemical potential with $m_{\rm L}=a\Lambda_{\rm QCD}$ for $0\leq a \leq 2$ and $m_{\rm T}=0$ for different values of $\Lambda_{\rm QCD}$. Though the critical chemical potential decreases as increasing the gluon mass $m_{\rm L}$, the gluon mass dependence is weaker  for large $m_{\rm L}$.

Further studies are needed for the mass function in entire range of the phase diagram, in order to know behaviors of the quark mass at strong coupling region.

\section*{Acknowledgements}

This work was partially supported by MEXT-Supported Program for
the Strategic Research Foundation at Private Universities, 2014-2017  
(S1411024).
 
%\newpage

%\vspace{5mm}

%\vspace{5mm}

%\begin{center}
%{\Large Appendix A}
%\end{center}
\vspace{5mm}

%-------------------------------------------------------------%                 %  REFERENCES
%%------------------------------------------------------------------------------%\newpage

\end{document}